\documentclass[aps,prx,twocolumn,superscriptaddress,longbibliography]{revtex4-1}
\usepackage[T1]{fontenc}
\usepackage[latin9]{inputenc}
\usepackage{graphicx,amssymb,amsmath,xcolor}
\usepackage[unicode=true,
 bookmarks=false,
 breaklinks=false,pdfborder={0 0 0},pdfborderstyle={},backref=false,colorlinks=true]
 {hyperref}
\hypersetup{
 colorlinks}


\begin{document}

\title{\textit{Ab initio} calculations of erbium crystal field splittings in oxide hosts: role of the $4f$ radial wave function}

\author{Yogendra Limbu}
\affiliation{Department of Physics and Astronomy, University of Iowa, Iowa City, Iowa 52242, USA}
\author{Yueguang Shi}
\affiliation{Department of Physics and Astronomy, University of Iowa, Iowa City, Iowa 52242, USA}
\author{Joseph Sink}
\affiliation{Department of Physics and Astronomy, University of Iowa, Iowa City, Iowa 52242, USA}
\author{Tharnier O. Puel}
\affiliation{Department of Physics and Astronomy, University of Iowa, Iowa City, Iowa 52242, USA}
\author{Durga Paudyal}
\affiliation{Department of Physics and Astronomy, University of Iowa, Iowa City, Iowa 52242, USA}
\affiliation{Ames National Laboratory of the US DOE, Iowa State University, Ames, IA 50011, USA}
\affiliation{Department of Electrical and Computer Engineering, Iowa State University, Ames, IA 50011, USA}
\author{Michael E. Flatt\'e}
\affiliation{Department of Physics and Astronomy, University of Iowa, Iowa City, Iowa 52242, USA}
\affiliation{Department of Applied Physics, Eindhoven University of Technology, Eindhoven, The Netherlands}

\begin{abstract}
\noindent We expand here our description of a newly developed simple and effective \textit{ab initio} method of calculating crystal field coefficients (CFCs) of rare-earth atoms in insulating hosts, focusing on Er in wide band gap oxide hosts MgO, ZnO, TiO$_2$, CaWO$_4$, and PbWO$_4$, which exemplify  different local site symmetries. These hybrid functional calculations, which incorporate a portion of exact exchange from Hartree-Fock theory, reproduce the experimentally identified insulating band gaps of these oxides.
The negative values of cohesive and formation energies confirm the structural and chemical stability of these oxides, whereas the defect formation energies indicate that Er doped ZnO, CaWO$_4$, and PbWO$_4$ are easier to form  compared to Er doped MgO and TiO$_2$. Er doped in these oxide hosts exhibits a  spin magnetic moment of $\sim$ 3 $\mu_B$ confirming the $3+$ valance state. The CFCs of these hosted Er are determined from charge densities and potentials obtained from non-spin-polarized calculations, involving a 4\textit{f} core approximation. These CFCs are subsequently used to solve an effective Hamiltonian and generate the crystal field splitting of Er 4\textit{f}. These calculated Er 4\textit{f} energy levels are in good agreement with  available experiments. %The optical transition between the lowest level of the first excited state and the lowest level of the ground state is $\sim$ 1.5 $\mu$m. This wavelength is within the telecommunication range, indicating that these Er hosted oxides are potential candidate materials for quantum telecommunication.
\end{abstract}

\maketitle

\section{Introduction}
\noindent The partially filled 4$f$ electrons of rare earth (RE) atoms doped in wide band gap oxides are well shielded from the host  electrons. This enables exceptional optical and spin properties that significantly improve the efficiency of information storage and transmission over long distances, assisting in the implementation of the quantum memories and transducters that are critical to quantum telecommunication~\cite{grant2024Optical, thiel2011rare, kolesov2012optical, stevenson2022erbium, zhong2019emerging,kunkel2015rare, Ji2024Nano,kenyon2002recent}. 
The crystal field splitting of 4$f$ states plays a crucial role in  these properties, and are determined by the crystal field coefficients (CFCs). These CFCs can be extracted from experiment through a fitting procedure by iteratively minimizing the energy difference between the fit and experimental energy levels~\cite{duan2007vacuum}. However, it is highly desirable to develop a method that can directly calculate CFCs without relying on fitting parameters, preferably using $\textit{ab-initio}$ calculations~\cite{novak2013crystal, ning2007density,dai1998description}, which provide accurate ground state densities for the interacting systems~\cite{sun2016accurate,jones2015density}. This work expands on a short recent publication\cite{YogendraCrystalPRL} that outlined this approach by providing extensive additional information for the range of materials considered there.

The properties of 4$f$ electrons in a solid, above the Kondo temperature,  can be described by an effective atomic
Hamiltonian, including a free ion interaction and
single-particle crystal field Hamiltonian~\cite{novak2013crystal, mollabashi2020crystal}. The parameters associated with the free ion Hamiltonian can be determined either through experimental fitting procedures or $\textit{ab-initio}$ calculations~\cite{10.1063/1.455853}. These effective atomic parameters are universal and apply to all compounds, even though the crystal field coefficients may vary from host to host~\cite{mulak2000effective}. The CFCs associated with the crystal field Hamiltonian are calculated by minimizing the energy difference between the experimentally observed energy levels and those obtained from the parameterized Hamiltonian model through a fitting procedure~\cite{duan2007vacuum}. However, this method has several limitations: (1) experimental data may be inadequate for calculating CFCs in low symmetry crystals ~\cite{ning2007density}, (2) there is a need to account for the polarization of the crystal field relative to the crystal axes~\cite{novak2013crystal}, (3) experimental data might be insufficient for magnetic and super-conducting systems~\cite{novak2013}, and (4) metallic systems can be problematic due to the lack of 4$f$ - 4$f$ optical transitions~\cite{novak2013crystal}. Several theoretical and semi-empirical Hamiltonians have been proposed to construct crystal field models, including the point charge model~\cite{hutchings1964point}, the superposition model~\cite{newman1989superposition}, and the overlap model~\cite{malta1982simple}. However, the accuracy of these models depends on the constituent terms, which themselves depend on CFCs~\cite{mollabashi2020crystal,mulak2000effective}. Further, the extensively investigated point charge model is known to be inadequate for calculating CFCs, as it tends to underestimate their values~\cite{mulak2019we}.

An alternative way is $\textit{ab-initio}$ approach, which provides more accurate values for CFCs~\cite{novak2013crystal, ning2007density}. The complexity arises from the 4$f$ electrons of RE, which are strongly localized and shielded by the 5$p$ and 6$s$ electrons and generally remain inactive in chemical bonding. However, they can be perturbed by the crystal field environment created by the surrounding atoms. With 4$f$ electrons in the valance, there could be hybridization of 4$f$ states with other states from the surrounding atoms, as well as self-interactions among 4$f$ electrons. These reduce the crystal field effect on the 4$f$ shell of RE, which should be avoided for the accurate calculations of CFCs. These  complexities are removed by placing the 4$f$ electrons in the core, which contribute only the spherical components to the total density of the system~\cite{novak2013crystal}. Further, the maximally-localized Wannier functions approach with 4$f$ electrons in the valence is used to determine an adjustable parameter that controls the hybridization of 4$f$ electrons with other states. However, the value of adjustable parameter is different  for different RE ions. Another method is the pseudo-potential method with core approximation~\cite{ning2007density}. 
With this approach, the tail of the 4$f$ electrons may unphysically hybridize with the density functional theory (DFT)-valence states, which need to be orthogonal to each other~\cite{novak2013crystal,ning2007density}.

Here, we report a simple and effective method of calculating the CFCs from DFT using a pseudo-potential method with a 4\textit{f} core approximation. Our approach is applied to Er doped wide band gap oxides, such as MgO, ZnO, and TiO$_2$ with $O_h$, $C_{3v}$, and $D_{2h}$ local site symmetries, respectively, as well as to CaWO$_4$ and PbWO$_4$ with $S_4$ symmetry. The negative values of cohesive and formation energies confirm the structural and chemical stability of these oxides, whereas the defect formation energies indicate that Er doped ZnO, CaWO$_4$, and PbWO$_4$ are easier to form than Er doped MgO and TiO$_2$. Er doped in these oxide hosts show a spin magnetic moment of $\sim$ 3 $\mu_B$, confirming the 3+ valence state. The CFCs of these hosted Er are determined from charge densities and potentials obtained from non-spin-polarized calculations, with the 4\textit{f} core approximation. The extracted CFCs are then used to solve an effective Hamiltonian and generate the 4$f$ splitting energy levels using  newly developed and available programs for the energy levels of lanthanides (qlanth ~\cite{qlanth} and lanthanide ~\cite{edvardsson2001atomic} codes). The calculated energy levels using both codes are in good agreement with the available experiments~\cite{stevenson2022erbium,wortman1971optical}, confirming the applicability of these methods to any material systems involving RE for determining CFCs.

\section{Theoretical Formulation and Computational Details}
\subsection{Theoretical Formulation}

The effective Hamiltonian for  the 4$f^N$ electronic configuration of an RE ion in the host material is written as:
\begin{equation}
H_{\text{eff}} = H_{\text{FI}} + V_{\text{CF}},
\end{equation}
where $H_{\text{FI}}$ and $H_{\text{CF}}$ represent the free ion and crystal field Hamiltonians ~\cite{novak2013crystal,novak2013}. The former with the two and three body operators in the configuration interactions is expressed as~\cite{10.1063/1.455853}:
\begin{multline}
H_{\text{FI}} = H_0 + \sum_{k=1}^{6} F^k f_k + \zeta_{4f} A_{SO} 
\quad + \alpha L(L+1) \\ + \beta G(G_2) + \gamma G(G_7) + \sum_{i} t_i T^i + \sum_{h} m_hM^h + \sum_{f} p_fP^f,
\label{H_FI}
\end{multline}
where $H_0$ is a spherically symmetric one-electron Hamiltonian, $F^k$ and $f_k$ ($k$ = 0, 2, 4, 6) are electrostatic integral and angular parts of the electrostatic interaction in the two-body configuration, $\zeta_{4f}$ and $A_{SO}$ are spin-orbit integral and angular parts of the spin-orbit interaction, $\alpha$, $\beta$, and $\gamma$ are the Trees parameters (correction terms, which are obtained from \textit{ab initio} or fitting experimental data~\cite{10.1063/1.455853}) associated with the two-body interaction operators~\cite{rajnak1963configuration, trees1951configuration}, $G(G_2$) and $G(G_7$) are the eigenvalues of Casimir's operators for the groups $G_2$ and $G_7$, and $L$ is the total orbital angular momentum~\cite{rajnak1963configuration,wybourne1965spectroscopic}. For  three  or more  4$f$ electrons, the three-particle configuration interaction is included by adding  $t_iT^i$ ($i$ = 2, 3, 4, 6, 7, 8) into the two particle Hamiltonian~\cite{judd1966three, rajnak1963configuration}, where $T^i$ denotes the adjustable
parameters, which are also obtained from an \textit{ab initio} method or from fitting experimental
data~\cite{10.1063/1.455853}, and $t_i$ are three-particle operators. Furthermore, $M_h$ ($h$ = 0, 2, 4)  are  spin-spin and spin-other-orbit relativistic corrections, known as  Marvin integrals~\cite{marvin1947mutual}, whereas $P^f$ ($f$ = 2, 4, 6) represents the electrostatically correlated spin-orbit perturbation of the two-body magnetic corrections.  $m_h$ and $p^f$ are operators associated with these magnetically correlated corrections. In our case, all the parameters associated with the free ion Hamiltonian are taken from Ref.~\cite{10.1063/1.455853}.

The crystal field Hamiltonian ($V_{\text{CF}}$) in Wybourne notation is written as:
\begin{equation}
V_{\text{CF}} =  \sum_{k, q} B_{q}^k (r) C_{q}^k (\theta, \phi),
\end{equation}
where $B_q^k$(r) and $C_{q}^k$($\theta$, $\phi$) are radial and spherical components of $V_{\text{CF}}$~\cite{wybourne1965spectroscopic}. $C_{q}^k$($\theta$ is expressed in terms of  spherical harmonics,
\begin{equation}
C_{q}^k (\theta, \phi) = \sqrt{\frac{4\pi}{2k+1}} Y_{q}^k (\theta, \phi).
\end{equation}
DFT provides the local potential, 
\begin{equation}
    V_{\text{Local}}(r) = V_{\text{I}}(\text{r}) + V_{H}(\text{r}) + V_{\text{XC}}(\text{r}),
\end{equation}
where V$_{I}$, V$_{H}$, and V$_{\text{XC}}$ represent ionic, Hartree, and exchange-correlation potentials, respectively. The Hartree and exchange-correlation potentials depend upon the electron densities. 
The local potential acts as a perturbation on the 4$f$ electrons of RE ions, making the spherically symmetric 4$f$ wave functions asymmetric. %(FIG.~\ref{symmetric_asymmetric_wavefunction}). 
This asymmetry is crucial for CFCs, which are calculated by taking the expectation value of $B^k_q(r)$ over the 4$f$ radial wave function.
\begin{equation}
\begin{aligned}
B^k_q &= \langle R_{4f}(r) | B^k_q(r) | R_{4f}(r) \rangle \\
&= \frac{2k+1}{4\pi} \int R^2_{4f}(r) V_{\text{CF}}(r) C^{*k}_q(\theta, \phi) r^2 \sin(\theta) \, dr \, d\theta \, d\phi.
\end{aligned}
\label{Bkq_eqa}
\end{equation}

Here we use an effective hydrogenic 4$f$ radial wave function. The spreading of the 4$f$ wave function (and thus the crystal field splittings)  are sensitive to the effective dielectric constant ($\epsilon_{\omega, q}$) of the medium near  the RE ion.  Nonetheless, the 4$f$ wave function typically extends up to a few \AA. The value of $\epsilon_{\omega, q}$ is different from the macroscopic value of the dielectric constant in the solid, which can be predicted by calculating crystal field energy levels at variable values of $\epsilon_{\omega, q}$ and compared with available experimental energy levels. However it is also different from the dielectric constant of vacuum. We express the 4$f$ radial wave function as
\begin{equation}
R_{4f} (r,\epsilon_{\omega,\mathbf{q}\sim a_0}) = A r^3 e^{-rZ_{\text{eff}}/{na_0\epsilon_{\omega,q}}},
\end{equation}
where  $A$, $n$, $a_0$, and $Z_{\text{eff}}$  are, respectively, the amplitude of the 4$f$ wave function, the principal quantum number, the Bohr radius, and the effective nuclear charge in the RE ion. The effective nuclear charge is calculated from
\begin{equation}
Z_{\text{eff}} = n\zeta_{n, l, m},
\end{equation}
where $\zeta_{n, l, m}$ is the orbital exponent of the 4$f$
electrons and depends on their screening constant, which is taken from Ref.~\cite{clementi1967atomic}. We used 27.9784 for the effective nuclear charge of erbium, which also controls the 4$f$ wave function in the host crystals. Then, the CFCs are calculated from  Eq.~(\ref{Bkq_eqa}) via numerical integration. To determine the value of $\epsilon_{\omega, q}$, the CFCs  at different values of $\epsilon_{\omega, q}$ are calculated, and these CFCs are then used to generate the  corresponding energy levels using qlanth and lanthanide codes~\cite{qlanth,edvardsson2001atomic}.  The calculated energy levels are then compared with the experiment. 

\subsection{Computational Details}

Density function theory (DFT)~\cite{sholl2022density} calculations are performed  to study the electronic structure and quantum properties of pristine and Er doped wide band gap oxides: MgO, ZnO, TiO$_2$, CaWO$_4$, and PbWO$_4$, using Vienna \textit{Ab-initio} Simulation Package~\cite{hafner2008ab,kresse1996efficiency}. Projector augmented wave (PAW) pseudo-potentials~\cite{blochl1994projector} and Perdew-Burke-Ernzerhof (PBE)~\cite{perdew1996generalized} functionals with gradient approximation (GGA)~\cite{perdew1986accurate} are used. In pseudo-potentials, 3$s^2$ for Mg, 2$s^2$ 2$p^4$ for O, 3$d^{10}$ 4$s^2$ for Zn, 3$d^3$ 4$s^1$ for Ti, 3$p^6$ 4$s^2$ for Ca, 5$d^5$ 6$s^1$ for W, 6$s^2$ 6$p^2$ for Pb, and 4$f^{11}$ 5$s^2$ 6$s^2$ 5$p^6$ 5$d^1$ for Er are considered as valance electrons for the electronic and magnetic properties calculations. However, the 4$f$ electrons are placed in core for the CFCs calculations. A sufficient plane wave cut off energy of 500 is utilized with energy threshold of 10$^{-6}$ eV. The geometries are relaxed until the force of each of the atoms becomes equal or less than 10$^{-5}$ eV/\AA.  We used a sufficient $k$ points grid of 10 $\times$ 10 $\times$ 10 in MgO, 8 $\times$ 8 $\times$ 5 in ZnO, 5 $\times$ 5 $\times$ 8 in TiO$_2$, and 4 $\times$ 4 $\times$ 2 in both CaWO$_4$ and PbWO$_4$ for the Brillouin zone sampling. The density of states (DOS) are calculated using Gaussian smearing method~\cite{jorgensen2021effectiveness} with smearing value of 0.10 eV.  

For the CFCs calculations, we made a sufficiently large super-cells: 3 $\times$ 3 $\times$ 3 containing 216 atoms (107 Mg and 107 O) for MgO,  4 $\times$ 4 $\times$ 3 containing 192 atom (96 Zn and 96 O) for ZnO, 3 $\times$ 3 $\times$ 4 containing 216 atoms (72 Ti and 144 O) for TiO$_2$, and 3 $\times$ 3 $\times$ 1 containing 216 atom (36 Ca/Pb, 36 W and 144 O) for both CaWO$_4$ and PbWO$_4$, which are enough to create the crystal field environment for the Er dopant. A single Er atom is then substituted at the Mg/Zn/Ti/Ca sites. The complexity arises  due to the strongly correlated 4$f$ electrons. When these electrons are taken as valance electrons, there is a hybridization of these 4$f$ states with other valance states (Fig.~\ref{Er_doped_pdos}) and also suffering from 4$f$ - 4$f$ onsite self interactions~\cite{novak2013crystal}, which results in reduction of the crystal field effect. The hybridization issue can be removed by applying Hubbard potential ($U$) on the top of GGA, however, it increases electronic complexity, which needs to be fixed for the optical transition of the 4$f$ states~\cite{novak2013}. One way to resolve this issue is to place the 4$f$ electrons in the core, which provides the spherical contribution to the total electron density~\cite{novak2013crystal}. The core treatment of 4$f$ electrons is possible for the augmented
plane waves basis~\cite{novak2013crystal}. Other methods require different method of treating the 4$f$ electrons~\cite{ning2007density}. Thus, we performed non-spin-polarized calculations with the 4$f$ electrons  treated as core electrons to generate the charge densities and local potentials. These are then used to extract the CFCs, leading to spin independent electron densities.

\begin{figure}[ht!]
\includegraphics[width=0.98\columnwidth]{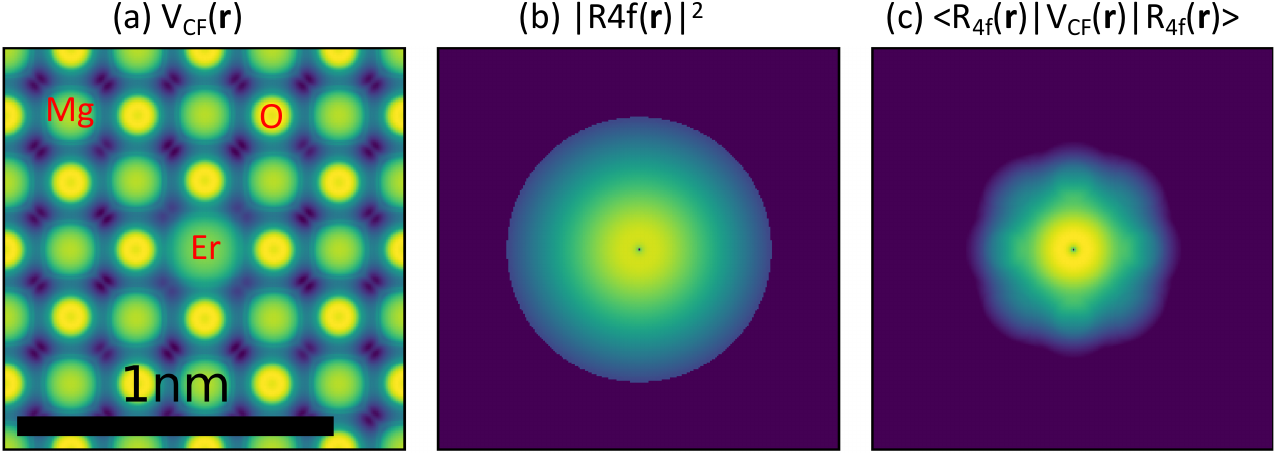}
\caption[dop]{(a) The DFT generated crystal field potential in Er doped MgO with a 216 atom supercell. (b) The magnitude squared of the bare $R_{4f}$(\textbf{r}) is spherically symmetric. However, the crystal field potential created by the surrounding atoms around the Er dopant introduces asymmetry in (c) the potential experienced by the Er 4$f$ states. This asymmetry is crucial for the calculation of CFCs.}
\label{symmetric_asymmetric_wavefunction}
\end{figure}

An example of the crystal field potential generated by DFT for Er in MgO is shown in Fig.~\ref{symmetric_asymmetric_wavefunction}(a). In Fig.~\ref{symmetric_asymmetric_wavefunction}(b) the 4$f$ radial wave function's magnitude squared is shown, followed by in (c) the overlap of the two that will produce the crystal field coefficients from Eq.~(\ref{Bkq_eqa}).

\begin{figure}[ht!]
\includegraphics[width=0.98\columnwidth]{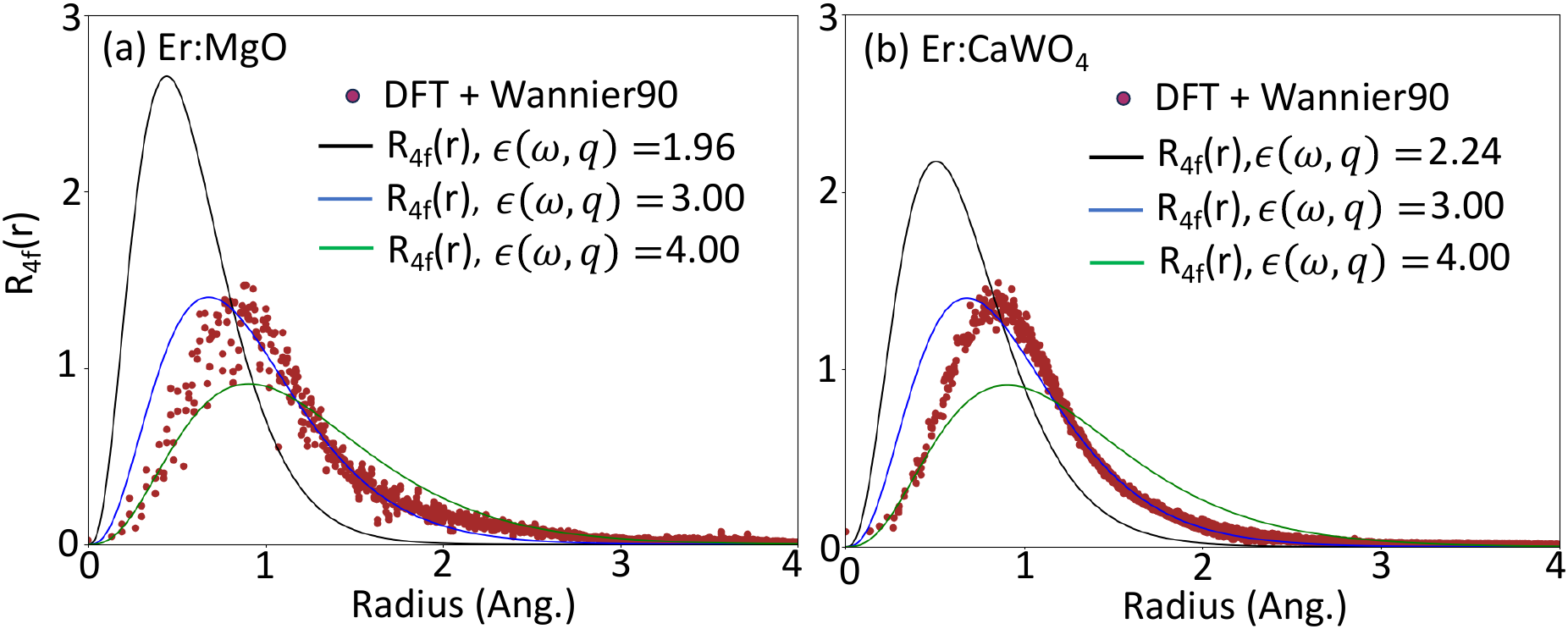}
\caption[dop]{Comparison of extracted 4$f$ wave functions from a maximally localized Wannier function (Wannier90) treatment with analytical solutions of hydrogenic 4$f$ radial wave functions in Er doped MgO (a) and Er doped CaWO$_4$ (b). The extracted 4$f$ radial wave functions compares well with the hydrogenic radial wave functions  when  $\epsilon_{\omega, q}$ is in between 3 and 4.}
\label{DFT_Wannier90}
\end{figure}

We find that a value of $\epsilon_{\omega, q}$  is close to 2 for wide band gap oxides of this study:  1.96 for  MgO, 1.90 ZnO, 2.00 for TiO$_2$, 2.24 for PbWO$_4$ and CaWO$_4$. These values yield energy levels that are in good agreement with experimental results. To assess this value we extract  the 4$f$ wave functions using the maximally localized Wannier functions  approach (Wannier90)~\cite{marzari2012maximally} and  compare them with the hydrogenic 4$f$ radial wave function form for different values of $\epsilon_{\omega, q}$ (FIG.~\ref{DFT_Wannier90}). The extracted 4$f$ wave functions align closely with the analytical solution of the hydrogenic radial wave function when $\epsilon_{\omega, q}$ is  between 3 and 4. Further, $\epsilon_{\omega, q}$ is also verified from the ionization energy of Er$^{3+}$ ($\sim$ 42.42 eV~\cite{johnson2017lanthanide, lang2010ionization}), which is expressed as, $E_{i}=\frac{m_e e^4}{(4\pi\epsilon_{\text{eff}})^2{\hbar^2}}\frac{Z_{\text{eff}}^2}{n^2}$, where  m$_e$ and  $\epsilon_{\text{eff}}$(= $\epsilon_{\omega, q}$$\epsilon_0$) are the mass of the electron and the permittivity. The calculated value of  $\epsilon_{\omega, q}$, is 3.96, which is in good agreement with the Wannier90 result ({FIG.~\ref{DFT_Wannier90}}). This confirms that $\epsilon_{\omega, q}$ is indeed small and is closer to the anticipated value ($\sim$ 2). For example, the comparison between the extracted 4$f$ radial wave functions of Er in MgO and  CaWO$_4$ from DFT + Wannier90 calculations and the analytical solution of the hydrogenic 4$f$ radial wave function at different values of $\epsilon_{\omega, q}$ are presented in FIG.~\ref{DFT_Wannier90}. For a much smaller value of $\epsilon_{\omega, q}$ (= 1), the 4$f$ radial wave function is localized below 2 \AA. For increased dielectric constant the wave function starts to spread out into the crystal, indicating that the wave function is sensitive to $\epsilon_{\omega, q}$.

\section{Results and Discussion}

\subsection{Structural Properties, and Phase Stability of Pristine and Er Doped Phases}
We studied five different oxides with distinct point and space group symmetries: MgO (cubic) with $O_h$ point and $Fm{\overline{3}m}$ space groups, ZnO (hexagonal) with $C_{6v}$ point and P6$_3$mc space groups, TiO$_2$ (rutile, tetragonal) with $D_{4h}$ point and P4$_2$/mnm space groups, and both CaWO$_4$ and PbWO$_4$ (tetragonal) with $C_{4h}$ point and I4$_1$/a space groups. The crystal structures and lattice constants of these host oxides are first optimized and then utilized in investigating their electronic structures properties. The optimized lattice parameters ($a$ = 4.24~\AA~ in MgO, $a$ =  3.28~\AA~ and $c$  = 5.30~\AA~ in ZnO, $a$ = 4.66~\AA and $c$ = 2.96~\AA~ in TiO$_2$, $a$ = 5.29~\AA~ and $c$ = 11.43~\AA~ in CaWO$_4$, and $a$ = 5.51~\AA~ and $c$ = 12.13~\AA~ in PbWO$_4$)
are in good agreement with available experiments~\cite{cimino1966dependence,karzel1996lattice,isaak1998elasticity,cavalcante2012electronic,isik2022effect}.
The structural and chemical stabilities of the pristine phases are investigated by calculating the cohesive and formation energies using formula mentioned in Ref.~\cite{limbu2022electronic}. The computed values of the formation (cohesive) energies are -2.722 (-5.185) in MgO, -1.434 (-3.693) in ZnO, -2.914 (-6.941) in TiO$_2$, -2.633 (-6.635) in CaWO$_4$, and -1.927 (-6.110) in PbWO$_4$. The negative values of these energies confirm the phase stability of these pristine oxides.
Similarly, the defect formation energy of the Er doped phases are calculated by employing:
$E_f= E_d^{Er}- E_p-E_{iso}(Mg/Zn/Ti/Ca/Pb) + E_{iso}(Er)$~\cite{limbu2022electronic}, where $E_d^{Er}$ and $E_p$ are the total energies with Er doped and pristine phases. $E_{iso}$(Mg/Zn/Ti/Ca/Pb) and $E_{iso}$(Er) are the isolated energies of Mg/Zn/Ti/Ca/Pb and Er atoms. The  calculations are performed with an Er doping concentration of 25\% (a single Er out of 4 Mg/Ti/Ca/Pb) in MgO, TiO$_2$, CaWO$_4$, and PbWO$_4$, while 12.5\% Er doping concentration  (a single Er out of 8  Zn) in ZnO. The calculated defect formation energies are 0.025 eV/atom in Er doped MgO, -0.312 eV/atom in Er doped ZnO, and  0.293 eV/atom in Er doped TiO$_2$, and -0.036 and -0.159 eV/atom in Er doped  CaWO$_4$ and PbWO$_4$. This defect formation energy analysis suggests that the formation of Er doped Zno, CaWO$_4$, and PbWO$_4$ is easier than that of Er doped MgO and TiO$_2$. We note that the values of defect formation energies in MgO and TiO$_2$ decrease with larger super cells: 0.007 eV/atom in MgO with a 2 $\times$ 2 $\times$ 2  supercell (3.125\% of Er: one Er out of 32 Mg), and 0.050 eV/atom in TiO$_2$ with a 2 $\times$ 2 $\times$ 3  supercell (4.17\% of Er: one Er out of 24 Ti).

\subsection{Electronic and Magnetic Properties of Undoped Phases}

\begin{figure*}[ht!]
\includegraphics[width=0.90\linewidth]{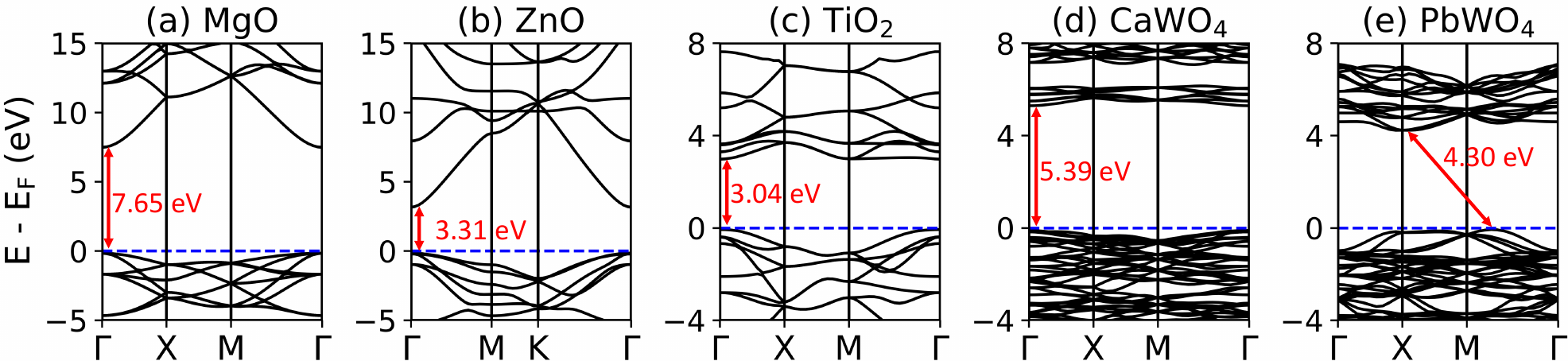}
\caption[dop]{Band structures  of undoped (a) MgO, (b) ZnO, (c) TiO$_2$, (d) CaWO$_4$, and (e) PbWO$_4$.}
\label{pristine_band}
\end{figure*}

Here, we describe the electronic and magnetic properties of undoped phases. GGA calculations show that all the studied oxides are  non-magnetic  with band gaps of 4.46 eV in MgO, 0.66 eV in ZnO, 1.66 eV in TiO$_2$, 3.95 eV in CaWO$_4$, and 3.25 eV in PbWO$_4$. These band gaps are smaller than available experimental values~\cite{whited1973exciton,srikant1998optical,serpone2006band,mikhailik2004one,lacomba2008optical}. However, the standard hybrid functional (HSE06) calculations correct these band gaps to 6.22 eV for MgO, 2.39 eV for ZnO,  3.04 eV for TiO$_2$, 5.39 eV for CaWO$_4$, and 4.30 eV for PbWO$_4$, which are in good agreement with the experimental values, except for MgO and ZnO~\cite{serpone2006band,mikhailik2004one,lacomba2008optical}(Fig.~\ref{pristine_band}). However, the use of Hartree Fock (HF) mixing parameters of 0.45 and 0.38 in hybrid functional (HSE) calculations correct the band gaps to 7.65 eV in MgO and 3.31 eV in ZnO. Thus, HSE corrected band gaps are in good agreement with the available experiments~\cite{whited1973exciton,srikant1998optical} (Fig.~\ref{pristine_band}). Additionally, all the studied oxides show direct band gaps, except for PbWO$_4$~\cite{roessler1967electronic,zhang1998electronic,das2014microscopic,kang2010quasiparticle}. Thus, these simulations of host oxides with wide band gaps are suitable for further treatment of RE ion doping~\cite{stevenson2022erbium}.

The partial density of states (PDOS) show that the conduction band is primarily contributed from the O-2$p$ states in MgO. However, in the unoccupied region, there is an almost equal contribution from O-2$p$ and Mg-3$s$ states. In ZnO, the top of the valence band is mainly coming from O-2$p$ states, with minor but not negligible contributions from Zn-3$d$ states. In the core region (-5.80 eV to -7.75 eV), however, there is a significant contribution from Zn-3$d$ states. In the conduction band, there are empty states coming from both O and Zn atoms, as in MgO. In TiO$_2$, the valance and conduction bands are mainly dominated from O-2$p$ and Ti-3$d$ states, respectively. The top of the valence band near the Fermi level is dominated by $p_x$  state of O, and the bottom of the conduction band is contributed from $d_{x^2-y^2}$ of Ti. Similarly, in CaWO$_4$, the conduction band is mainly contributed from O-2$p$ states, and small but not negligible contribution from W-5$d$ and Ca-3$p$ states. The conduction band is mainly formed from the W-5$d$ states.  The Ca atom shows a relatively minor contribution to both valence and conduction bands as compared to other atoms. A similar characteristic is found in PbWO$_4$ as in CaWO$_4$, with the notable distinction of a significant contribution from Pb-6$s$ states in the energy range of -7.2 eV to -8.1 eV.

\subsection{Electronic Properties of Er Doped Material}
\begin{figure}[ht!]
\includegraphics[width=0.98\columnwidth]{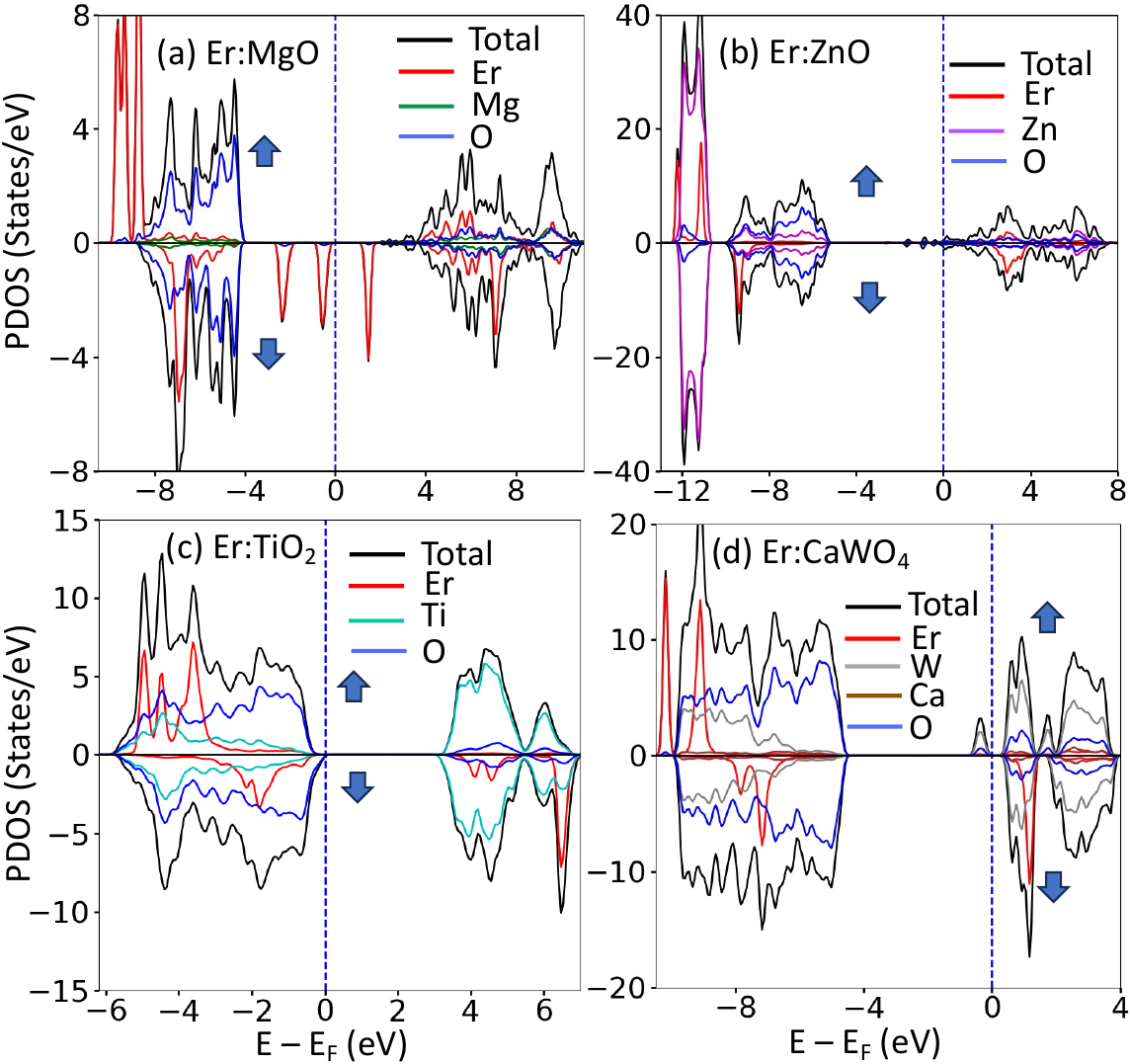}
\caption[dop]{Partial density of states (PDOS) of Er doped (a) MgO, (b) ZnO,  (c) TiO$_2$, and (d) CaWO$_4$ with HSE calculations.}
\label{Er_doped_pdos}
\end{figure}

For the electronic and magnetic properties of Er doped oxides, we use a 25\% doping concentration in all cases except for ZnO, where a 12.5\% doping concentration is used. In Er doped MgO, there are two distinct 4$f$ density of states (DOS) peaks located below the Fermi level and one distinct 4$f$ DOS peak located above the Fermi level in the gap. In addition, the well separated 4$f$ states are found in the region between -8.50 eV to -10 eV. Thus, only the 4$f$ states spreading from -4.35 to -8.50 eV are hybridized with the oxygen atom states, which results in less crystal field effect on 4$f$ states of Er. The calculated spin magnetic moment of Er 4\textit{f} is 2.92 $\mu_B$ ($\sim$ 3 $\mu_B$), indicating that the Er has ${3+}$ oxidation state. The 4$f$ states do appear to be strongly hybridized with the Zn-3$d$ states in Er doped ZnO. In the case of Er doped TiO$_2$, the 4$f$ states are mainly hybridized with O-2$p$ states (Fig.~\ref{Er_doped_pdos} (c)). In both Er doped CaWO$_4$ and PbWO$_4$, the 4$f$ states exhibit similar characteristics. A single, well-separated 4\textit{f} state is found around -10 eV. Between -6 eV and -10 eV, the 4$f$ states are hybridized with the O and W atoms. The contributions from Ca and Pb atoms to the PDOS are negligible, as shown in Fig.~\ref{Er_doped_pdos} (d) for Ca. Thus, the crystal field splitting of Er in these oxides is primarily caused by the O-2$p$ states and the local site symmetry of the RE ion in the host crystals.

\subsection{Crystal Field Coefficients (CFCs)}
\begin{figure}[ht!]
\includegraphics[width=0.99\columnwidth]{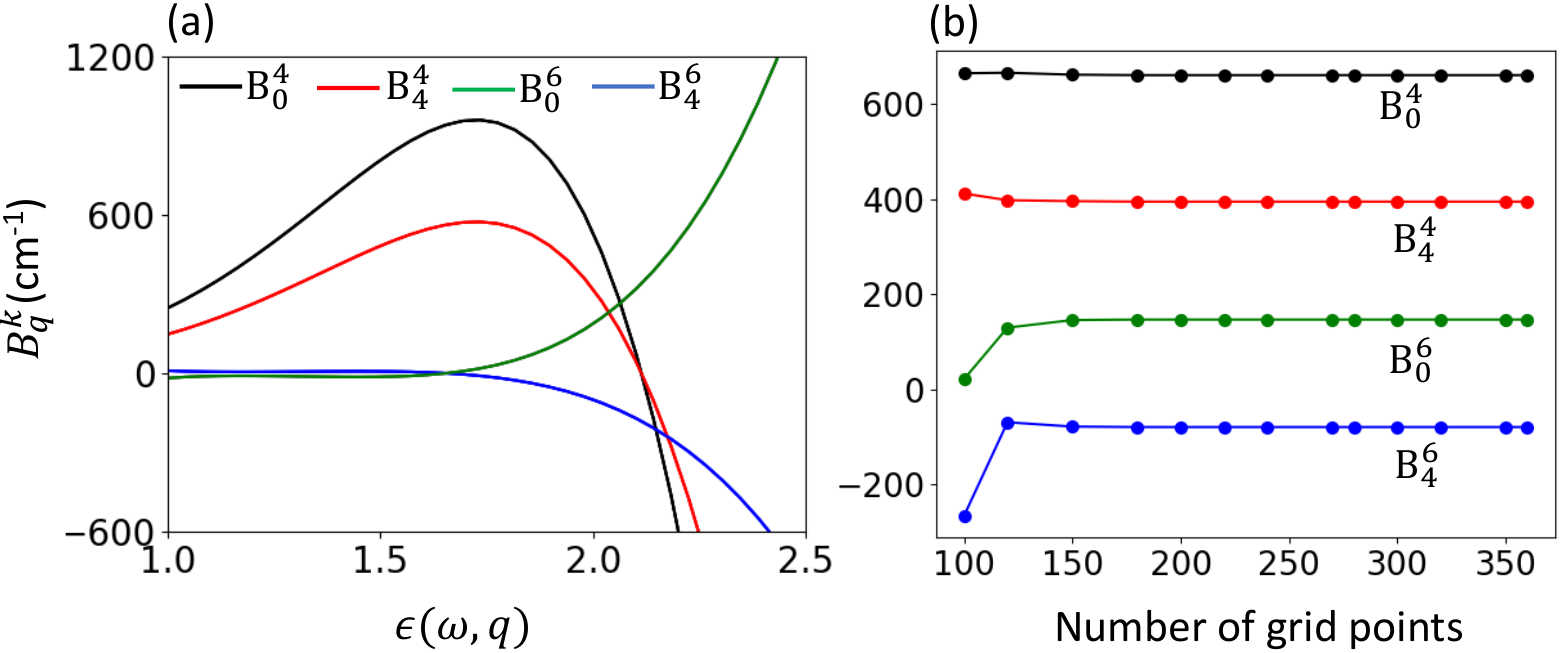}
\caption[dop]{CFCs of Er doped MgO as a function of $\epsilon (\omega,q)$ (a) and the magnitude of non-zero CFCs at different number of grid points (b).}
\label{Bkq_N_E}
\end{figure}

The Er atom occupies the Mg site in MgO with O$_h$ site symmetry. Due to the high symmetry of its crystal structure, only two independent non-zero CFCs ($B^4_0$ and $B^6_0$) are needed to describe the crystal field splitting of the 4$f$ states of Er. The remaining CFCs are calculated by using the relations {$B_{\pm{4}}^4$} = $\sqrt{\frac{5}{14}}B_{0}^4$ and  {$B_{\pm{4}}^6$} = -$\sqrt{\frac{7}{2}}{B_{0}^6}$~\cite{ning2007density}. Here, we use Wybourne notation and units of cm$^{-1}$~\cite{wybourne1965spectroscopic}. The computed CFCs are shown in Table~\ref{CFCs_table}. The value of $B^6_0$ is smaller as compared to $B^4_0$, consistent with Cs$_2$NaYbF$_6$ with $O_h$ symmetry~\cite{ning2007density}. The accuracy of calculations depends upon the numerical integration of Eq.~(\ref{Bkq_eqa}), which relies on a fine grid mesh. To test the accuracy, multiple calculations were performed using different grid meshes, ranging from 100 $\times$ 100 $\times$ 100 to 400 $\times$ 400 $\times$ 400 for the cubic crystal of MgO, with 50 $\times$ 50 $\times$ 50 increment in each crystallographic directions, as shown in Fig.~\ref{Bkq_N_E}. We find that the values of non-zero CFCs stabilize after 150 grid points in each direction, confirming that this grid size is sufficient for accurately determining CFCs. For a lower symmetry structure, there are many non-zero CFCs  as shown in Table~\ref{CFCs_table}, depending on the local site symmetry of the RE ions.

 Er occupies the Zn site with $C_{3v}$ local symmetry in ZnO. There are six non-zero CFCs: $B^2_0$, $B^4_0$, $B^4_{\pm3}$, $B^6_0$, $B^6_{\pm3}$, and $B^6_{\pm6}$, as described in Ref~\cite{divivs2008crystal}. $B^4_{\pm3}$ and $B^6_{\pm3}$ have only imaginary components, whereas the remaining CFCs have only real components  (Table~\ref{CFCs_table}). In CaWO$_4$/PbWO$_4$, Er occupies the Ca/Pb site with $S_4$ symmetry. There are five non-zero CFCs, where only $B^4_{\pm4}$ and $B^6_{\pm4}$ have both real and imaginary components, as mentioned in Ref.~\cite{mollabashi2020crystal}. The value of $B^2_0$ is relatively larger as compared to the other non-zero values. Similarly, in TiO$_2$, Er is substituted at Ti site with $D_{2h}$ local symmetry, which is lower  than the other symmetries ($O_h$, $C_{3v}$, and $S_4$). Thus, there are  nine non-zero CFCs: $B^2_0$, $B^2_{\pm2}$, $B^4_0$, $B^4_{\pm2}$, $B^4_{\pm4}$, $B^6_0$, $B^6_{\pm2}$, $B^6_{\pm4}$, and $B^6_{\pm6}$, as mentioned in Ref.~\cite{furrer1988neutron}.  For $q$ = 2 and 6, only imaginary components are present, whereas the remaining CFCs have only real components. $B^4_0$ and $B^6_0$ have opposite signs and are larger in magnitude compared to the other CFCs.

\begin{table}[ht!]
\hfill{}
\begin{tabular}{l c c c c c}
\toprule
$B_q^k$ & MgO   & ZnO  & TiO$_2$  & CaWO$_4$  & PbWO$_4$ \\
\hline
$B_{0}^2$ &    &  -260.96  & 74.19 & 846.34  & 872.91\\
$B_{\pm{2}}^2$ &  & & $\pm$$i$474.97 &  &  \\
$B_{0}^4$ & 660.96 & 199.04 & -493.49 & 187.43  & 90.08\\
$B_{\pm{2}}^4$ &   &   & $\pm$$i$146.80 &  &   \\
$B_{\pm{3}}^4$ &   &  -$i$92.07 & &  &   \\
$B_{\pm{4}}^4$ &  395.00  &  & 310.25 & 
$\left(
\begin{array}{c}
     192.40\\
     \mp i56.58
\end{array}
\right )$  & 
$\left ( 
\begin{array}{c}
114.55\\
\pm i2.79
\end{array}
\right )$
\\
$B_{0}^6$ & -78.57  &  -240.17 & 466.69 &  87.40 & 138.50\\
$B_{\pm{2}}^6$ &   &  & $\pm$$i$85.65 &  &  \\
$B_{\pm{3}}^6$ &   & -$i$228.92 &  &  &  \\
$B_{\pm{4}}^6$ & 146.98 &  & 106.82 & 
$\left ( \begin{array}{c}
     311.06\\
     \mp i424.36
\end{array} \right)$   
& $\left ( \begin{array}{c}
     182.07\\
     \mp i352.72
\end{array}\right)$\\
$B_{\pm{6}}^6$ &  & 233.60 & $\pm$$i$254.93 &   & \\
\hline
\hline
\end{tabular}
\hfill{}
\caption{Table shows the extracted CFCs of Er doped in wide band gap oxides with variable local site symmetry, e.g., MgO with $O_4$, ZnO with $C_{3v}$, TiO$_2$ with $D_{2h}$, and CaWO$_4$ and PbWO$_4$ with $S_4$, from DFT calculations. All the CFCs are expressed in cm$^{-1}$. }
\label{CFCs_table}	
\end{table}

\begin{figure*}[ht!]
\includegraphics[width=0.80\linewidth]{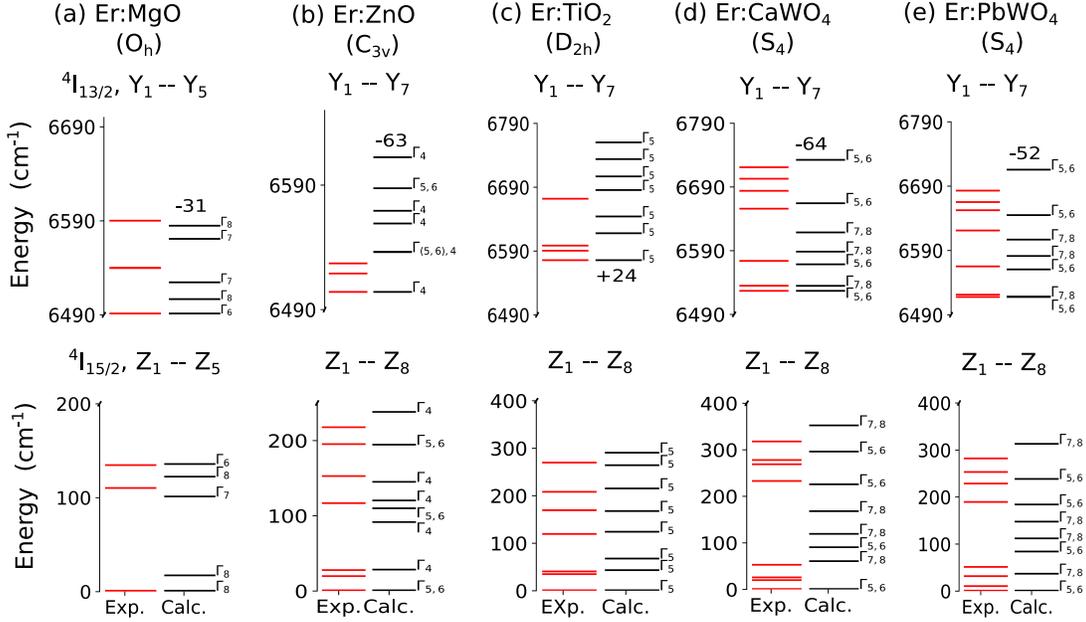}
\caption[dop]{The ground and first excited state energy splitting of 4$f$ states of Er$^{3+}$ ions in (a) MgO, (b) ZnO, (c) TiO$_2$, (d) CaWO$_4$, and (e) PbWO$_4$. The eight doubly degenerate energy levels due to Kramers degeneracy, $Z_1$ - $Z_8$, in the ground state and seven energy levels,
$Y_1$ - $Y_7$, in the first excited state of Er$^{3+}$ are identified in all cases except for MgO, where  five energy levels are found in both the ground and first excited states.}
\label{Er_splitting}
\end{figure*}

\subsection{Crystal Field Splitting Energies}

%Embedding a single Er ion in host materials, its symmetry is reduced, leading to the splitting of its 4$f$ states. This splitting of the 4$f$ states depends on the site symmetry of the Er ion within the host material. 
Er has a total orbital angular momentum ($L$) of 6 and a total spin angular momentum ($S$) of 3/2, resulting in a total angular momentum ($J$) of 15/2. Thus, in the free Er$^{3+}$ ion, there are 52  degenerate energy levels, calculated as (2$S$ + 1)(2$L$ + 1). The spin-orbit coupling (SOC) splits these degenerate energy levels into sixteen levels in the ground state ${}^4$I$_\frac{15}{2}$ and fourteen levels in the first excited state ${}^4$I$_\frac{13}{2}$, and so on. Further, the crystal field effect splits these degenerate energy levels into doubly or quarterly degenerate multiplets, depending on the site symmetry of Er.

In Er doped MgO, Er has $O_h$ symmetry. The ground state splits into three quartets $\Gamma_8$ and two doublets $\Gamma_7$ and $\Gamma_6$, as also suggested in Refs.~\cite{bleaney1959new, borg1970spin} and observed experimentally ~\cite{stevenson2022erbium}. Similarly, two quartets and  three doublets are found in the first excited state, which are in good agreement with the experiment~\cite{stevenson2022erbium}. The ground and first excited states are represented by $Z_i$ (where $i$ = 1, 2, 3, 4, and 5) and $Y_j$ (where $j$ = 1, 2, 3, 4, and 5), which are labelled by irreducible representation ($\Gamma$), as mentioned in Ref.~\cite{bradley1976p}. The optical transition from $Y_1$ ($\Gamma_6$) to $Z_1$ ($\Gamma_8$) is 6522.43 cm$^{-1}$ ($\sim$ 1.53 $\mu$m), which corresponds to the telecommunication C-band~\cite{nishi2005optical}. 

In Er doped ZnO with $C_{3v}$, TiO$_2$ with $D_{2h}$, CaWO$_4$ with $S_4$, and PbWO$_4$ with $S_4$, there are eight Kramers doublets in the ground state and seven Kramers doublets in the first excited state, which are represented by $Z_i$ (where $i$ = 1, 2, 3, 4, 5, 6, 7, and 8) and $Y_j$ (where $j$ = 1, 2, 3, 4, 5, 6, and 7) in the ground and first excited states. In Er doped ZnO, the ground state multiplets are represented  by three  $\Gamma_{5,6}$ and five $\Gamma_4$  irreducible representations, whereas there are two $\Gamma_{5,6}$ and five $\Gamma_4$ in the first excited state. The optical transition from $Y_1$ to $Z_1$ is  6567.52 cm$^{-1}$, which is in good agreement with experiment~\cite{stevenson2022erbium}. However, in the case of Er doped TiO$_2$, the ground and first excited states multiplets are represented by a single irreducible representation of $\Gamma_5$, as mentioned in Ref.~\cite{goodman1991crystal}. The 4$f$ - 4$f$ transition from $Y_1$ to $Z_1$ is  6552.43 cm$^{-1}$, which is (23.27 cm$^{-1}$) smaller than the experiment~\cite{stevenson2022erbium}. In addition,  the eight  (4$\Gamma_{5,6}$ + 4$\Gamma_{7,8}$) in the ground states and seven  (4$\Gamma_{5,6}$ + 3$\Gamma_{7,8}$) Kramers doublets in the first excited states are identified in Er doped in both CaWO$_4$ and PbWO$_4$ due to the same $S_4$ symmetry. The transitions from $Y_1$ to $Z_1$ is   6591.30 cm$^{-1}$, which is 64.30 cm$^{-1}$ larger than the experiment~\cite{wortman1971optical} in Er doped CaWO$_4$, while 6570.05 cm$^{-1}$ in Er doped PbWO$_4$, showing a good agreement with experiment~\cite{stevenson2022erbium}.

\section{Conclusion}
We have developed an effective \textit{ab initio} method of calculating crystal field coefficients of Er$^{3+}$ ions in wide band gap oxides with different local site symmetries, such as MgO with $O_h$, ZnO with $C_{3v}$, TiO$_2$ with $D_{2h}$, and both CaWO$_4$ and PbWO$_4$ with $S_4$. The hybrid functional calculations reproduce the insulating experimental band gap of these oxides suitable for the 4$f$ optical transitions of RE ions. The negative values of cohesive and formation energies confirm the structural and chemical stability of these oxides, while the defect formation energies indicate that Er doped ZnO, CaWO$_4$, and PbWO$_4$ are easier to form as compared to Er doped MgO and TiO$_2$. Er doped in these oxide hosts show  spin magnetic moment of $\sim$ 3 $\mu_B$ confirming the 3+ valance state. The CFCs are computed from the charge densities and potentials generated by non spin polarized DFT calculations using 4\textit{f} core approximation. The extracted CFCs are then used to solve an effective Hamiltonian and generate
the crystal field splitting of 4$f$ states of Er, which  are in good agreement with available experiments. We identify eight, $Z_1$ - $Z_8$, and seven, $Y_1$ - $Y_7$, doubly degenerate ground and first excited state energy levels of Er$^{3+}$ ions in ZnO, CaWO$_4$ and PbWO$_4$, and TiO$_2$. However, Er$^{3+}$ in MgO shows three quartets  and two doublets in the ground state and two quartets and three doublets in the first excited state. The  calculated 4$f$ intra-band  transition  from the lowest energy level of the first excited state to the lowest energy level of the ground state, $Y_1$ $\rightarrow$ $Z_1$, is $\sim$ 1.5 $\mu$m, which  is within the telecommunication range, thereby making these oxides potential candidate materials for quantum telecommunication.

\section*{Acknowledgment}
This work is supported by the U.S. Department of Energy, Office of Science, Office of Basic Energy Sciences under Award Number DE-SC0023393.

\end{document}